\documentclass{article}

\usepackage{arxiv}

\usepackage[utf8]{inputenc}
\usepackage[T1]{fontenc}
\usepackage{hyperref}
\usepackage{url}
\usepackage{booktabs}
\usepackage{amsfonts}
\usepackage{nicefrac}
\usepackage{microtype}
\usepackage{lipsum}
\usepackage{graphicx}
\usepackage{natbib}
\usepackage{amsmath}
\usepackage{amssymb} 
\usetikzlibrary{calc}
\usepackage{doi}
\usepackage{float}

\title{Morse-based Modular Homology for Evolving Simplicial Complexes}

\date{August 17, 2025}

\author{ \href{https://orcid.org/0009-0004-8771-3616}{\includegraphics[scale=0.06]{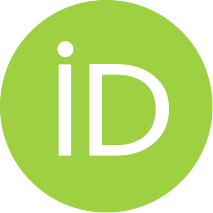}\hspace{1mm}Anqiao Ouyang$^{1}$} \\
        Mount Boucherie Secondary School \\
        West Kelowna, British Columbia, Canada \\\\
	Department of Aerospace Engineering \\
	San Diego State University\\
	San Diego, California, USA \\
	\texttt{aouyang@sdsu.edu} \\
}
\renewcommand{\shorttitle}{Morse-based Modular Homology for Evolving Simplicial Complexes}

\hypersetup{
pdftitle={Morse-based Modular Homology for Evolving Simplicial Complexes},
pdfsubject={},
pdfauthor={Anqiao Ouyang},
pdfkeywords={},
}

\begin{document}
\maketitle
\begin{abstract}
    The computation of homology groups for evolving simplicial complexes often requires repeated reconstruction of boundary operators, resulting in prohibitive costs for large-scale or frequently updated data. This work introduces MMHM, a Morse-based Modular Homology Maintenance framework that preserves homological invariants under local complex modifications. An initial discrete Morse reduction produces a critical cell complex chain-homotopy equivalent to the input; subsequent edits trigger localized updates to the affected part of the reduced boundary operators over a chosen coefficient ring. By restricting recomputation to the affected critical cells and applying localized matrix reductions, the approach achieves significant amortized performance gains while guaranteeing homology preservation. A periodic recompression policy together with topology-aware gating and a column-oriented sparse boundary representation with a pivot-ownership map confines elimination to the affected columns and can bypass linear algebra when invariants are decidable combinatorially. The framework offers a drop-in upgrade for topology pipelines, turning costly rebuilds into fast, exact updates that track homology through local edits. Reframing dynamic homology as a locality-bounded maintenance task provides an exact alternative to global recomputation for evolving meshes and complexes.

\end{abstract}


\section{Introduction}

Topological invariants play a fundamental role in the analysis and classification of geometric structures.  
In the context of computational topology, the homology groups of a simplicial complex capture essential global features such as connected components, independent cycles, and voids, while remaining insensitive to continuous deformations \citep{edelsbrunner2002topological, hatcher2002algebraic}.  
These invariants are particularly critical in low-dimensional settings, dimensions zero through three, where they serve as compact descriptors in geometry processing, scientific simulation, and topological data analysis (TDA) \citep{carlsson2009topology, zomorodian2005computing}.  
Efficient computation and maintenance of homological information have become increasingly important in applications where the underlying domain undergoes frequent local modifications.

Traditional homology computation relies on full reduction of the boundary matrices derived from the simplicial complex \citep{munkres1984elements}.  
Although robust, such static methods are computationally expensive when applied repeatedly in dynamic settings, where the topology of the domain changes through local operations such as edge flips, local refinement or coarsening, and hole creation or closure. In many practical scenarios, only a small region of the complex is affected by each update, yet the full recomputation cost remains proportional to the size of the entire reduced complex.  
This mismatch between update locality and computational cost motivates the development of algorithms capable of maintaining homology incrementally with complexity depending primarily on the size of the affected region \citep{bauer2021ripser}.

The challenge is compounded when integer coefficients are required, as in the detection of torsion subgroups.  
In such cases, the computation of Smith normal form (SNF) can dominate the runtime if applied to full matrices after each update. An ideal dynamic framework would limit both reduction and torsion verification to the relevant submatrices while preserving correctness guarantees.

\subsection{Topological Invariants in Low Dimensions}

For a finite simplicial complex $K$, the $k$-th homology group over a coefficient ring $R$ is defined as
\[
H_k(K; R) \;=\; \ker \partial_k \,/\, \operatorname{im} \partial_{k+1}
\]
where $\partial_k$ denotes the $k$-th boundary operator mapping $k$-chains to $(k-1)$-chains.  When $R = \mathbb{F}$ is a field, $H_k(K; \mathbb{F})$ is a finite-dimensional vector space whose dimension $\beta_k$ is the $k$-th Betti number, counting the number of independent $k$-dimensional holes.  
When $R = \mathbb{Z}$, the structure theorem for finitely generated abelian groups yields the decomposition
\[
H_k(K; \mathbb{Z}) \;\cong\; \mathbb{Z}^{\beta_k} \;\oplus\; \bigoplus_{i=1}^t \mathbb{Z}_{p_i}
\]
where $p_i > 1$ are the torsion coefficients. These coefficients reveal finite-order homology classes, which arise naturally in spaces such as lens spaces and certain manifold triangulations.

In low dimensions ($0 \leq k \leq 3$), Betti numbers and torsion factors admit a direct geometric interpretation: $\beta_0$ counts connected components, $\beta_1$ counts independent loops or tunnels, and $\beta_2$ counts enclosed voids, with torsion encoding additional finite-order relations. This interpretability has made low-dimensional homology a core descriptor in geometry processing, scientific simulation, and topological data analysis (TDA).

In many applications, the underlying simplicial complex evolves through localized modifications while preserving or subtly altering topology.  
Typical operations include edge flips in two dimensions, local refinement and coarsening in adaptive meshes, Pachner moves in three dimensions \citep{pachner1991pl}, and topological events such as hole creation or closure through the insertion or deletion of minimal simplex sets.  
Although these updates typically affect only a star-shaped neighborhood of limited size $s$, traditional homology algorithms often require recomputation over the entire reduced complex, incurring costs proportional to the global number of critical simplices rather than to $s$.  
This disparity between update locality and computational expense motivates the design of dynamic frameworks that limit reduction and torsion verification to the affected region, while ensuring correctness guarantees for both field and integer coefficients.

\subsection{Combinatorial Representation of Simplicial Complexes}

A finite simplicial complex $K$ provides a combinatorial abstraction of a geometric domain, representing both its topology and discrete geometry through a set of simplices glued along shared faces. Formally, a $k$-simplex $\sigma^k = [v_0, v_1, \dots, v_k]$ is the convex hull of $k+1$ affinely independent vertices, and the faces of $\sigma^k$ are obtained by deleting one or more vertices. A simplicial complex is then defined as a finite collection of simplices closed under the face relation and with the intersection of any two simplices being a face of each.  

Such a combinatorial representation admits a purely algebraic encoding via chain groups and boundary operators \citep{hatcher2002algebraic}. For each dimension $k$, the $k$-chain group $C_k(K; R)$ is the free $R$-module generated by the oriented $k$-simplices of $K$. The boundary operator $\partial_k : C_k \to C_{k-1}$ is defined on an oriented simplex by  
\[
\partial_k [v_0, v_1, \dots, v_k]  
= \sum_{i=0}^k (-1)^i [v_0, \dots, \widehat{v_i}, \dots, v_k]
\]  
and extended linearly to $C_k(K; R)$. This algebraic--combinatorial framework provides a sparse, structured representation of topology that can be directly translated into matrices over $R$, serving as the basis for algorithmic computation of homology groups.

In computational topology, the boundary matrix representation not only enables efficient reduction-based algorithms \citep{zomorodian2005computing} but also facilitates integration with combinatorial optimizations such as discrete Morse theory \citep{forman2002user}. By storing simplices in dimension-wise order and indexing them consistently, one obtains a compact, machine-friendly encoding that supports both static and dynamic updates to the complex. This makes simplicial complexes a natural choice for homology computation in settings where the underlying space evolves through localized combinatorial modifications.

\subsection{Related Work}
A foundational approach to dynamic homology over $\mathbb{Z}_2$ is to rebuild all boundary matrices and perform a complete reduction after each modification. This guarantees correctness for coefficients but scales with the size of the reduced complex, making it impractical for highly localized updates. The following paragraphs situate full recomputation and coreduction within the literature, and contrast them with persistence-based and localized viewpoints.

\paragraph{Full recomputation.}
Reconstructing and fully reducing the boundary matrices after every edit yields correct Betti numbers for arbitrary update sequences. Standard matrix-reduction–based pipelines, as surveyed in computational topology, have cubic worst-case algebraic cost in the number of cells and can be memory-intensive on large instances \citep{edelsbrunner2010computational}. While straightforward and robust, this strategy discards prior eliminations and ignores the typically small support of local edits, leading to poor scaling when updates are frequent and localized.

\paragraph{Coreduction (discrete Morse cancellations).}
Coreduction applies discrete Morse cancellations to remove reducible face–coface pairs before any algebraic reduction, shrinking the chain complex while preserving homology \citep{forman2002user,MrozekBatko2009Coreduction}. As a preprocessing step, coreduction can dramatically reduce boundary-matrix sizes, often down to the critical cells, so that subsequent linear algebra runs on a much smaller problem \citep{DlotkoEtAl2011CoreductionCW,HarkerEtAl2013FoCM}. This principle underlies efficient pipelines in both plain and persistent settings \citep{MrozekWanner2010CoreductionInclusionsPH}. In favorable instances (e.g., large contractible regions or acyclic substructures), the speedups are substantial; in unfriendly cases with few cancellable pairs, the benefit is limited \citep{HarkerEtAl2013FoCM,mischaikow2013morse}. As a per-update baseline, a “coreduction-only’’ procedure re-runs cancellations after each edit and then fully reduces the residual matrix, so the overall cost may still be dominated by global operations under frequent updates.

\paragraph{Persistence-based reductions (static/semi-dynamic).}
Persistent homology algorithms perform a single low-pivot elimination across a filtration, from which plain homology of any level can be read \citep{zomorodian2005computing}. Treating a sequence of insertions as a filtration amortizes work across stages and often runs much faster in practice than repeatedly solving from scratch, despite cubic worst-case bounds. However, classical persistence assumes monotone growth; handling deletions requires additional machinery and can diminish computational advantages in fully dynamic settings. Discrete Morse ideas have also been used alongside persistence to reduce fill-in and matrix sizes prior to elimination \citep{lewiner2003optimal,mischaikow2013morse}.

\paragraph{Localized/dynamic viewpoints.}
Beyond global recomputation, localized strategies rebuild and reduce only the subcomplex affected by an edit, while accounting for how that region attaches to the remainder of the complex. Such ideas leverage the small support of local updates but must still perform linear algebra on the impacted submatrix and can lose efficacy when edits influence large-scale cycles. Modern hybrid frameworks integrate Morse-theoretic compression with modular (localized) linear algebra so that only a small critical submatrix is updated after each edit; ablations that disable either the Morse compression or the localization clarify the trade-off between matrix size and update scope in low-dimensional settings.

\section{Method}

\subsection{Problem Formulation}

Let $K = \bigcup_{k=0}^d K_k$ be a finite simplicial complex of dimension $d \leq 3$, where
\[
K_k = \{\sigma^k_1, \dots, \sigma^k_{n_k}\}
\]
denotes the set of $k$-simplices. Over the coefficient field $\mathbb{F}=\mathbb{Z}_2$, the simplicial chain groups
\[
C_k(K;\mathbb{F}) \cong \mathbb{F}^{n_k}
\]
form a chain complex
\[
\cdots \xrightarrow{\partial_{k+1}} C_k \xrightarrow{\partial_k} C_{k-1} \xrightarrow{\partial_{k-1}} \cdots \xrightarrow{\partial_1} C_0 \to 0,
\]
where $\partial_k : C_k \to C_{k-1}$ are the boundary operators satisfying $\partial_k \circ \partial_{k+1} = 0$.

In the dynamic setting, $K$ evolves under a sequence of local update operations, each consisting of insertions or deletions of simplices within the \emph{star} of a small set of simplices. The objective is to maintain the homology groups
\[
H_k(K;\mathbb{F}) \;=\; \ker \partial_k \big/ \operatorname{im} \partial_{k+1}
\]
and the corresponding Betti numbers
\[
\beta_k \;=\; \dim_{\mathbb{F}} H_k(K;\mathbb{F}),
\]
without recomputing all boundary matrices from scratch after each update.

Formally, if $K_t$ denotes the complex after $t$ updates and $\mathbf{B}_k^{(t)}$ denotes the matrix representation of $\partial_k$ at time $t$ with respect to a fixed ordering of simplices, the goal is to update
\[
\big(\operatorname{rank}\,\mathbf{B}_1^{(t)}, \dots, \operatorname{rank}\,\mathbf{B}_{d}^{(t)} \big)
\]

hence $\beta_0^{(t)}, \dots, \beta_d^{(t)}$—in time that depends primarily on a locality parameter $s \ll |K_t|$ (e.g., polynomial in $s$) rather than on the total complex size $|K_t|$. Here $s$ may be taken as the total number of simplices across all dimensions whose incidences change during the update. Note that $\partial_0 \equiv 0$, so $\mathbf{B}_0^{(t)}$ is the zero matrix and $\operatorname{rank}\,\mathbf{B}_0^{(t)}=0$.

\subsection{Boundary Matrix Representation}

Let $K$ be a finite simplicial complex of dimension $d \leq 3$. For each $0 \leq k \leq d$, the $k$-chain group over a coefficient field $\mathbb{F}$ is
\[
C_k(K;\mathbb{F}) \cong \mathbb{F}^{n_k},
\]
with the canonical basis given by the oriented $k$-simplices of $K$. The $k$-th boundary operator
\[
\partial_k : C_k(K;\mathbb{F}) \longrightarrow C_{k-1}(K;\mathbb{F})
\]
is defined on a basis element $\sigma = [v_0, v_1, \dots, v_k]$ by
\[
\partial_k \sigma = \sum_{i=0}^k (-1)^i [v_0, \dots, \widehat{v_i}, \dots, v_k],
\]
where the hat denotes omission. The matrix representation $B_k$ of $\partial_k$ with respect to the canonical bases has one column per $k$-simplex and one row per $(k{-}1)$-simplex; entries are $+1$ or $-1$ according to induced orientation, and $0$ otherwise. Over $\mathbb{Z}_2$, signs are omitted so entries are in $\{0,1\}$. Note that $\partial_0 \equiv 0$, hence $B_0$ is the zero matrix and $\operatorname{rank}(B_0)=0$.

In the implementation, each $B_k$ is stored in a sparse \emph{column-oriented} (CSC) layout to support efficient column replacement under local updates. All simplices are normalized to an ordered tuple of vertex indices before insertion to ensure a canonical representation up to orientation. A \emph{low-pivot map} is maintained in the \emph{row $\to$ column} convention: for each row index $r$, $\mathrm{low}(r)=j$ records that $r$ is the pivot (lowest nonzero entry) of column $j$ in the reduced form. This is equivalent to the common column $\to$ row mapping in persistent-homology implementations, but the chosen direction facilitates localized pivot-ownership tracking and conflict resolution.

Given the ranks $r_k = \operatorname{rank}(B_k)$, the Betti numbers satisfy
\[
\beta_0 = n_0 - r_1,\qquad
\beta_k = n_k - r_k - r_{k+1}\ \ (1 \le k \le d-1),\qquad
\beta_d = n_d - r_d,
\]
where $n_k=\dim C_k(K;\mathbb{F})$ is the number of $k$-simplices. The same mapping applies when working on a critical complex $K^C$ (obtained via discrete Morse reduction): replace $B_k$ by the corresponding critical boundary matrix $B_k^C$ and $n_k$ by the number of critical $k$-simplices $n_k^C$.

Forman's discrete Morse theory is used to shrink the chain complex while preserving homology \citep{forman2002user}. A discrete Morse matching on $K$ is a set
\[
\mathcal{V} \;\subset\; \bigcup_{k=1}^d \bigl\{\,(\tau^{k-1},\sigma^{k}) \;\big|\; \tau^{k-1}\prec \sigma^{k}\,\bigr\}
\]
such that (i) each simplex of $K$ appears in at most one pair of $\mathcal{V}$, and (ii) the directed graph induced by the matching contains no closed $V$-paths\footnote{A $V$-path (a gradient path) is an alternating sequence 
$\tau_0 \prec \sigma_0 \succ \tau_1 \prec \sigma_1 \succ \cdots \prec \sigma_{m-1} \succ \tau_m$
with $(\tau_i,\sigma_i)\in\mathcal{V}$ for all $i$, where each $\tau_{i+1}$ is a codimension-$1$ face of $\sigma_i$ distinct from $\tau_i$. 
A matching is valid iff it contains no closed $V$-path; equivalently, the induced gradient vector field is acyclic.}. 
Here $\tau^{k-1}\prec \sigma^{k}$ denotes that $\tau$ is a codimension-$1$ face of $\sigma$.

Let $S_k^C \subseteq K_k$ be the set of \emph{critical} $k$-simplices (those not in any pair of $\mathcal{V}$). Define the \emph{critical chain groups}
\[
C_k^C(K;\mathbb{F}) \;:=\; \mathrm{span}_{\mathbb{F}}\bigl(S_k^C\bigr),\qquad 0\le k\le d,
\]
and write
\[
K^C \;:=\; \bigoplus_{k=0}^d C_k^C.
\]
The boundary operators
\[
\partial_k^C: C_k^C \longrightarrow C_{k-1}^C
\]
are induced by the matching via $V$-paths (Forman’s gradient flow) and assemble into the \emph{critical chain complex}
\[
0 \longrightarrow C_d^C \xrightarrow{\;\partial_d^C\;} C_{d-1}^C \longrightarrow \cdots \longrightarrow C_0^C \longrightarrow 0,
\]
which is chain-homotopy equivalent to the original complex. Consequently,
\[
H_\ast(K;\mathbb{F}) \;\cong\; H_\ast(K^C;\mathbb{F}).
\]
In particular, homology can be computed from the \emph{critical boundary matrices} $B_k^C$ (the matrix representations of $\partial_k^C$ with respect to the canonical bases of $C_{k-1}^C$ and $C_k^C$). Here, “reduction’’ refers to Morse cancellations and should not be confused with row/column-reduced linear-algebraic forms.

In the present framework, discrete Morse reduction is applied at initialization to obtain $K^C$ from $K$, typically yielding $|S_k^C|\ll |K_k|$ and thereby smaller boundary matrices. During dynamic updates, a \emph{periodic recompression policy} maintains these size and sparsity advantages: the critical complex is rebuilt at fixed step intervals and whenever the fraction of affected simplices in an update exceeds a prescribed threshold. This prevents drift in the size of $K^C$ and keeps the critical boundary matrices, as well as any topological-gating counters, synchronized with the evolving complex.

\paragraph{Periodic recompression policy}
Let $t\in\mathbb{N}$ index the updates and let $m\in\mathbb{N}$ and $\tau\in(0,1)$ be user-chosen parameters.
Write $C_k^C(t)$ for the critical $k$-chains before applying update $t$ and let $R_k^C(t)\subseteq C_k^C(t)$ be the set of critical $k$-simplices whose column incidences change due to update $t$.
Define the locality ratio
\[
\rho_t \;:=\; \frac{\sum_{k=0}^d |R_k^C(t)|}{\max\!\bigl(1,\ \sum_{k=0}^d |C_k^C(t)|\bigr)}
\]
A \emph{recompression event} is triggered at update $t$ if any of the following holds:
\begin{enumerate}\itemsep2pt
  \item \textbf{Periodic trigger:} $t \equiv 0 \pmod m$.
  \item \textbf{Locality trigger:} $\rho_t \ge \tau$.
  \item \textbf{Validity trigger:} a gating precondition fails (e.g., a non-manifold edge with $\deg(e)\notin\{1,2\}$ is detected, or a manifold/connectedness assumption required by a gate is violated).
\end{enumerate}
When a recompression event fires, a fresh discrete Morse matching is computed on $K_t$ to obtain new critical sets $S_k^C(t{+}1)$; the critical boundary matrices $B_k^C(t{+}1)$ are rebuilt, the pivot-ownership map is re-initialized, topological-gating counters (such as edge-incidence tallies) are synchronized, and the ranks $r_k$ are recomputed for $k=1,\dots,d$.
In the absence of a recompression event, only the localized reducer operates on the affected columns $\{\ B_k^C[:,j]\ :\ j\in R_k^C(t)\ \}$ and any columns with pivot conflicts.
Typical choices are $m=32$ and $\tau=0.30$; these yield near-flat per-update latency with isolated spikes at recompression steps.

\subsection{Localized Boundary Updates}

Consider a local modification of the simplicial complex $K$ involving the insertion or deletion of simplices within the \emph{star} of a small set of simplices.
For each dimension $k$, let
\[
R_k \subseteq K_k
\]
denote the set of $k$-simplices whose incidences in the boundary operators are affected by the modification.
When $|R_k| \ll |K_k|$, it is advantageous to update only the corresponding \emph{columns} of the boundary matrices rather than recomputing them in their entirety.

In the context of a critical complex $K^C$ obtained from discrete Morse reduction, the affected simplices induce corresponding index sets
\[
R_k^C = \{\, j \mid \sigma_j^k \in C_k^C \ \text{and} \ \sigma_j^k \in R_k \,\}.
\]
Let $\mathbf{B}_k^C$ denote the $k$-th \emph{critical} boundary matrix.
Updating the homology then reduces to replacing the column supports
\[
\{\, \mathbf{B}_k^C[:, j] \mid j \in R_k^C \,\}
\]
with their new incidence patterns, followed by reapplying column-reduction operations only to these affected columns and to any columns whose pivots conflict with them.
If the update inserts or deletes $(k{-}1)$-simplices (thus changing the \emph{row} set of $\mathbf{B}_k^C$), the pivot map of $\mathbf{B}_k^C$ is re-initialized for that dimension; whether to escalate to a full recompression is decided by the periodic recompression policy.

Affected index sets $R_k$ are computed via a face-difference procedure that compares the pre-update and post-update stars of the modified region.
All simplices are first normalized by sorting their vertex indices before insertion into the matrix (over $\mathbb{Z}_2$ the sign is irrelevant), ensuring a canonical ordering that eliminates spurious orientation or permutation mismatches.
The matrices $\mathbf{B}_k^C$ are stored in a \emph{column-oriented} (CSC) sparse layout, supporting efficient column replacement.

For boundary matrices over $\mathbb{F}=\mathbb{Z}_2$, a \emph{low-pivot map} is maintained
\[
\mathrm{low}: \{\text{row indices}\} \longrightarrow \{\text{column indices}\},
\]
where $\mathrm{low}(i)=j$ indicates that row $i$ is the pivot of column $j$ in the column-reduced form (this row$\to$column convention is equivalent to the more common column$\to$row mapping and is chosen to facilitate localized pivot-conflict resolution).
When a column is replaced, its pivot assignment is cleared, the column is re-reduced against existing pivots, and any resulting pivot conflicts are propagated in a localized manner.
This ensures that the rank of $\mathbf{B}_k^C$ is updated exactly, without reprocessing unaffected columns.

\subsection{Topological Gating}

In certain restricted classes of simplicial complexes, specific Betti numbers can be inferred directly from simple combinatorial invariants without performing matrix reduction.
The MMHM framework incorporates such topological gating to avoid unnecessary linear-algebraic computations when the value of a target Betti number can be determined \emph{a priori} from update-local information.
If a gating precondition is violated at any point, the algorithm reverts to the appropriate localized or full reduction mode.

\paragraph{$\beta_2$ gate (boundary-edge criterion over $\mathbb{Z}_2$).}
Consider a connected triangulated $2$-manifold.
Over $\mathbb{Z}_2$, one has $\beta_2=1$ if and only if the surface is \emph{closed} (i.e., has no boundary); if the surface has boundary, then $\beta_2=0$.
Let
\[
E_{\partial} \;=\; \{\, e \in K_1 \mid \deg(e)=1 \,\},
\]
where $\deg(e)$ is the number of incident $2$-simplices.
Under the manifold assumption (every interior edge has $\deg(e)=2$), this yields the exact criterion
\[
\beta_2 \;=\;
\begin{cases}
1, & \text{if } |E_{\partial}| = 0,\\[2pt]
0, & \text{if } |E_{\partial}| > 0.
\end{cases}
\]
If a non-manifold edge with $\deg(e)\notin\{1,2\}$ is detected, the gate is disabled.

\paragraph{Incremental maintenance of $|E_{\partial}|$.}
To enable constant-time evaluation after each update, maintain an edge-incidence map
\[
\delta: K_1 \longrightarrow \{0,1,2\},
\]
where $\delta(e)$ counts the number of incident $2$-simplices (truncated at $2$).
Upon insertion or deletion of a $2$-simplex $[u,v,w]$, update the degrees of its three edges in $O(1)$ time, and adjust
\[
|E_{\partial}| \;=\; \#\{\, e \in K_1 \mid \delta(e)=1 \,\}.
\]
This yields an updated $\beta_2$ value without invoking any boundary matrix operations.

When connectedness of the complex is guaranteed \emph{a priori}, $\beta_0$ is set directly to $1$ without any reduction.
For orientable $2$-manifolds, an optional $\beta_1$ gate can be applied using a primal–dual tree–cotree test to detect changes in the first homology class before engaging matrix reduction; in genus-$0$ surfaces with $L$ boundary components, one may compute $\beta_1 = \max(0,\,L-1)$ exactly.
In situations where only $\beta_2$ changes according to the boundary-edge count and all gating preconditions are satisfied, the update bypasses rank computations entirely, maintaining only the relevant combinatorial counters and recording metrics without performing any linear-algebraic operations.

\subsection{Betti Number Maintenance}

Let $\mathbf{B}_k^C$ denote the $k$-th boundary matrix of the critical complex $K^C$ obtained after discrete Morse reduction, with
\[
\mathbf{B}_k^C \in \mathbb{F}^{\,n_{k-1}^C \times n_k^C}, \qquad \mathbb{F}=\mathbb{Z}_2.
\]
By rank–nullity,
\[
\dim_{\mathbb{F}} \ker \mathbf{B}_k^C \;=\; n_k^C - \mathrm{rank}\,\mathbf{B}_k^C.
\]
Since
\[
\beta_k \;=\; \dim_{\mathbb{F}} \ker \mathbf{B}_k^C \;-\; \mathrm{rank}\,\mathbf{B}_{k+1}^C,
\]
computing all Betti numbers reduces to determining the ranks of the critical boundary matrices. In particular,
\[
\beta_0 = n_0^C - r_1,\qquad
\beta_k = n_k^C - r_k - r_{k+1}\ \ (1\le k \le d-1),\qquad
\beta_d = n_d^C - r_d,
\]
where $r_k := \mathrm{rank}\,\mathbf{B}_k^C$. (Note that $\partial_0 \equiv 0$, so $\mathrm{rank}\,\mathbf{B}_0^C=0$ and need not be maintained.)

To avoid full recomputation after each update, the framework caches
\[
\bigl(r_1,\dots,r_d\bigr) \quad \text{together with} \quad (n_0^C,\dots,n_d^C).
\]
When an update affects only a subset $R_k^C$ of the critical $k$-simplices, ranks are updated \emph{selectively}:
\begin{itemize}\itemsep2pt
  \item recompute $r_k$ if either the \emph{column set} of $\mathbf{B}_k^C$ changes ($R_k^C\neq\varnothing$) \emph{or} the \emph{row set} of $\mathbf{B}_k^C$ changes (equivalently, $R_{k-1}^C\neq\varnothing$);
  \item ensure that for each $k$, the pair $(r_k,r_{k+1})$ used in $\beta_k$ is refreshed whenever $R_{k-1}^C$, $R_k^C$, or $R_{k+1}^C$ is nonempty.
\end{itemize}
Ranks for unaffected dimensions are reused from the cache.

If a localized reducer is available in dimension $k^\ast$, the rank $r_{k^\ast}$ is updated incrementally by reassigning pivots in the \emph{column-reduced} form (over $\mathbb{Z}_2$), avoiding any operation on unaffected columns. This guarantees exact Betti numbers while ensuring that the computational cost scales with the size of the affected region rather than with the total complex size.

When the zeroth Betti number is known \emph{a priori} due to topological assumptions (e.g., connectedness), $\beta_0$ is set directly and no operation is performed in dimension $0$.
When a topological gate determines a target Betti number (e.g., $\beta_2$ via the boundary-edge criterion), the corresponding rank updates are bypassed.
A rebuild of the critical complex (periodic recompression) is triggered if the shape of any $\mathbf{B}_k^C$ changes substantially, or if a gating precondition is violated.

\subsection{Computational Complexity}

Let $|K|=\sum_{k=0}^d n_k$ denote the total number of simplices in the complex, and let $|C|=\sum_{k=0}^d n_k^C$ denote the total number of critical simplices after discrete Morse reduction. For a local update affecting $s_k$ simplices of dimension $k$ and $s=\max_k s_k$, the computational cost of different stages can be analyzed as follows.

Constructing all boundary matrices from scratch requires
\[
T_{\mathrm{assemble}}=\Theta\!\Big(\sum_{k=1}^d n_k\cdot (k{+}1)\Big)=\Theta(|K|),
\]
since each $k$-simplex has exactly $k{+}1$ codimension-$1$ faces. Restricting the assembly to the affected region of size $s$ reduces the cost to $\Theta(s)$.

Given the boundary matrices, a full discrete Morse reduction via coreduction operates in
\[
T_{\mathrm{Morse}} = O\!\Big(|K|+\sum_{k=1}^d \mathrm{nnz}(\mathbf{B}_k)\Big),
\]
i.e., linear in the number of incidences. When applied only to the affected region, this becomes $O\!\big(s+\mathrm{nnz}_{\mathrm{local}}\big)$, where $\mathrm{nnz}_{\mathrm{local}}$ counts nonzeros in the restricted submatrices.

For a critical boundary matrix $\mathbf{B}_k^C\in\mathbb{F}^{\,n_{k-1}^C\times n_k^C}$, let
$N_k:=\max\{n_{k-1}^C,n_k^C\}$. Sparse elimination over $\mathbb{Z}_2$ can suffer fill-in, giving
\[
\text{worst case:}\quad O(N_k^3)\qquad\text{(dense-like behavior)}.
\]
In typical sparse regimes one observes near–quadratic scaling,
\[
\text{typical:}\quad \tilde O(N_k^2)
\]
where $\tilde O(\cdot)$ suppresses mild polylog/implementation factors.

For a \emph{localized} update that touches $s_k^C$ critical $k$-simplices (i.e., $s_k^C$ columns in $\mathbf{B}_k^C$), column reduction can be confined to those columns and their pivot-conflict neighborhood; under sparse operations with limited fill-in,
\[
T_{\mathrm{reduce}}^{\mathrm{local}}
\;=\;
\Theta\!\Big(\sum_{k=1}^d (s_k^C)^2\Big)
\]
Writing $s^C:=\max_k s_k^C$ and using $d\le 3$ as a constant,
\[
T_{\mathrm{reduce}}^{\mathrm{local}}
\;=\;
\Theta\!\big((s^C)^2\big)
\]

\paragraph{Periodic recompression overhead.}
A periodic recompression policy rebuilds the critical complex at fixed step intervals and when the fraction of affected cells exceeds a threshold. A rebuild from the current $K$ costs
\[
T_{\mathrm{rebuild}}=\Theta\!\Big(|K|+\sum_{k=1}^d \mathrm{nnz}(\mathbf{B}_k)\Big)
\]
covering reassembly and Morse reduction. If the interval is $M$ updates, the steady-state amortized overhead contributes $T_{\mathrm{rebuild}}/M$ per update; occasional threshold- or validity-triggered rebuilds do not change the asymptotic scaling when they are infrequent. Rebuilds are also triggered when the shape of some $\mathbf{B}_k^C$ changes substantially or when a gating precondition is violated.

In regimes where the critical size remains bounded (e.g., $n_k^C$ and the effective sparsity patterns $\mathrm{nnz}(\mathbf{B}_k^C)$ do not drift significantly), the implicit constants in the $\tilde{O}(|C|^2)$ and $\Theta\!\big(\sum_k (s_k^C)^2\big)$ expressions remain stable, and the per-update complexity is empirically flat between recompressions, with isolated spikes at rebuild steps.

\section{Experiments}

\subsection{Baselines}\label{sec:baselines}
All baseline methods operate over $\mathbb{Z}_2$, adopt identical simplex orientations and orderings, and are evaluated on exactly the same update sequences (including batch size $B$ and locality radius/size $s$) to ensure fairness. An update refers to a local insertion/deletion in a chosen star, a $2$–$2$ edge flip (2D), or a Pachner move (3D). Unless otherwise noted, each baseline recomputes the Betti numbers from the corresponding boundary operators of the current complex state.

\textbf{Full-Recompute.} After every update, the entire set of boundary matrices is rebuilt and fully reduced over $\mathbb{Z}_2$ to obtain the ranks and hence the Betti numbers. This classical strategy is conceptually simple and robust to arbitrary edit sequences, but it discards all previously computed structure and repeatedly performs elimination on unaffected portions of the complex. The worst-case cost is cubic in the global matrix size, with practical performance governed by sparsity and fill-in; under frequent localized edits, this cumulative cost becomes prohibitive.

\textbf{Coreduction-only.} Discrete Morse–style coreductions are first applied to cancel reducible face–coface pairs, yielding a smaller chain complex of critical cells that preserves homology; a full reduction is then performed on the residual boundary matrices. This two-stage procedure, coreduction followed by elimination, is repeated from scratch after each update. When many cells are collapsible, the residual matrices are dramatically smaller and the elimination stage is inexpensive, leading to substantial speedups in practice. The advantage diminishes in instances with few reducible pairs, and without incremental maintenance of the cancellation structure, repeated scans can rediscover the same collapses across steps. Still, as a dynamic baseline, Coreduction-only captures the benefit of shrinking matrices before algebraic reduction.

\textbf{Static-PH Reduction (P1/P2 baseline).}
“Static-PH Reduction” is a baseline introduced in this work. It applies the standard persistent-homology low-pivot reduction \citep{zomorodian2005computing,edelsbrunner2002topological} to each snapshot in a \emph{static}, per-state manner: a filtration-compatible ordering is imposed and a single $\mathbb{Z}_2$ reduction is executed, after which plain homology is read from the terminal filtration index. This design leverages sparsity and the induced pivot structure, it serves as a strong static comparator. Accordingly, it is used for P1–P2, where per-state solvers are sensible.

\subsection{P1: A Minimal Topological Perturbation on Genus-0 Triangulations}

Let \(K_0\) be a 2D simplicial complex homeomorphic to \(\mathbb{S}^2\), initialized as the boundary of a regular octahedron with
\[
|V|=6,\quad |E|=12,\quad |F|=8
\]
Fix a window patch \(P\subset F\) consisting of two adjacent triangular 2-simplices that share an interior edge \(e\).

\medskip
\[
\textsc{Open}(P):\quad K \mapsto K \setminus \big(P \cup \{e\}\big),
\qquad
\textsc{Close}(P):\quad K \mapsto K \cup \{e\} \cup P
\]

\begin{figure*}[!hb]
\centering
\begin{tikzpicture}[scale=1, every node/.style={font=\small}]

  \def\A{2.0}
  \def\H{2.6}

  \begin{scope}[shift={(0,0)}]
    \coordinate (v1) at (-\A,-\A);
    \coordinate (v2) at ( \A,-\A);
    \coordinate (v3) at ( \A, \A);
    \coordinate (v4) at (-\A, \A);
    \coordinate (t)  at (0,\H);
    \coordinate (b)  at (0,-\H);

    \fill[faceP] (t) -- (v1) -- (v2) -- cycle;
    \fill[faceP] (t) -- (v2) -- (v3) -- cycle;
    \fill[face]  (t) -- (v3) -- (v4) -- cycle;
    \fill[face]  (t) -- (v4) -- (v1) -- cycle;
    \fill[face]  (b) -- (v2) -- (v1) -- cycle;
    \fill[face]  (b) -- (v3) -- (v2) -- cycle;
    \fill[face]  (b) -- (v4) -- (v3) -- cycle;
    \fill[face]  (b) -- (v1) -- (v4) -- cycle;

    \draw[edge] (v1)--(v2)--(v3)--(v4)--(v1);
    \foreach \x in {v1,v2,v3,v4} { \draw[edge] (t)--(\x); \draw[edge] (b)--(\x); }

    \foreach \p in {v1,v2,v3,v4,t,b}{ \node[vertex] at (\p) {}; }

    \node[anchor=north] at (0,-\H-1.0) {\textbf{Closed:} $\beta_2=1$, $|E_{\partial}|=0$};
  \end{scope}

  \begin{scope}[shift={(8.0,0)}]
      \coordinate (v1) at (-\A,-\A);
      \coordinate (v2) at ( \A,-\A);
      \coordinate (v3) at ( \A, \A);
      \coordinate (v4) at (-\A, \A);
      \coordinate (t)  at (0,\H);
      \coordinate (b)  at (0,-\H);
    
      \fill[face] (t) -- (v3) -- (v4) -- cycle;
      \fill[face] (t) -- (v4) -- (v1) -- cycle;
      \fill[face] (b) -- (v2) -- (v1) -- cycle;
      \fill[face] (b) -- (v3) -- (v2) -- cycle;
      \fill[face] (b) -- (v4) -- (v3) -- cycle;
      \fill[face] (b) -- (v1) -- (v4) -- cycle;
    
      \draw[edge] (v1)--(v2)--(v3)--(v4)--(v1);
      \draw[edge] (t)--(v1);
      \draw[edge] (t)--(v3);
      \draw[edge] (t)--(v4);
      \foreach \x in {v1,v2,v3,v4} { \draw[edge] (b)--(\x); }
    
      \draw[bedge] (t)--(v1);
      \draw[bedge] (v1)--(v2);
      \draw[bedge] (v2)--(v3);
      \draw[bedge] (t)--(v3);
    
      \foreach \p in {v1,v2,v3,v4,t,b}{ \node[vertex] at (\p) {}; }
    
      \node[anchor=north] at (0,-\H-1.0)
        {\textbf{Open:} $\beta_2=0$, $|E_{\partial}|=4$, $|E|=11$};
    \end{scope}
\end{tikzpicture}
\caption{Two-triangle window patch $P$ on a triangulated sphere (octahedron model)}
\label{fig:tikz_close_open}

\end{figure*}

Throughout the sequence the complex remains a connected genus-0 surface. 
On \textsc{Open} the two triangles of the patch \(P\) \emph{and their shared edge} are removed; this preserves manifoldness and creates exactly one boundary component. 
Consequently the vertex set is fixed (\(|V|=6\)), while the combinatorial counts toggle as
\[
|E|: 12 \leftrightarrow 11,\qquad |F|: 8 \leftrightarrow 6,
\]
corresponding to the closed and open states, respectively.

\medskip
\noindent The homological effect is completely determined by the boundary count \(b\in\{0,1\}\) on a genus-0 surface:
\[
\beta_0 = 1,\qquad \beta_1 = 0,\qquad 
\beta_2 = \begin{cases}
1,& b=0 \ (\text{closed}),\\
0,& b=1 \ (\text{open}).
\end{cases}
\]
Thus \(\beta_0\equiv 1\) and \(\beta_1\equiv 0\) are invariant, while \(\beta_2\) toggles between \(1\) (closed) and \(0\) (open).

Only \(\beta_2\) changes under this minimal local edit, so the benchmark isolates the per-update latency of homology maintenance from large-scale recomputation, with the combinatorial deltas restricted to \(\Delta |E|=\pm 1\) and \(\Delta |F|=\pm 2\) and a fixed vertex set.

\begin{table}[!hb]
\centering
\caption{Mean per-update runtime on P1}
\label{tab:p1_latency}
\begin{tabular}{lccc}
\toprule
Method & Amortized (ms/update) & Mean step time (ms) & Throughput (updates/s)\\
\midrule
MMHM        & 0.200 & 0.150 & 5000.0 \\
Full        & 0.585 & 0.356 & 1709.4 \\
Static--PH  & 0.604 & 0.371 & 1655.6 \\
Coreduction & 1.188 & 0.643 & 841.8 \\
\bottomrule
\end{tabular}
\end{table}

Table~\ref{tab:p1_latency} reports mean per-update latency and throughput on P1. 
MMHM achieves the lowest latency and highest throughput under this localized, minimal topological perturbation. 
In amortized latency, MMHM is \(2.93\times\) faster than Full-Recompute, \(3.02\times\) faster than Static--PH, and \(5.94\times\) faster than Coreduction. 
On mean step time, the gains are \(2.37\times\), \(2.47\times\), and \(4.29\times\), respectively. These improvements align with the design goal of restricting elimination to a local critical block while avoiding global fill-in.

\begin{figure}[H]
\centering
\begin{subfigure}[b]{0.47\textwidth}
    \centering
    \includegraphics[width=\textwidth]{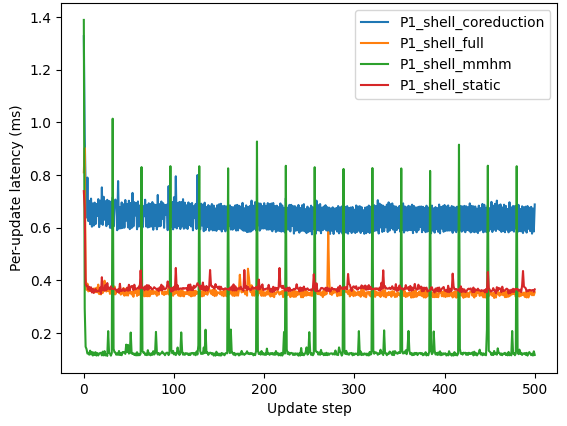}
    \caption{Per-update latency for P1 benchmark}
    \label{fig:p1_latency}
\end{subfigure}
\hfill
\begin{subfigure}[b]{0.5\textwidth}
    \centering
    \includegraphics[width=\textwidth]{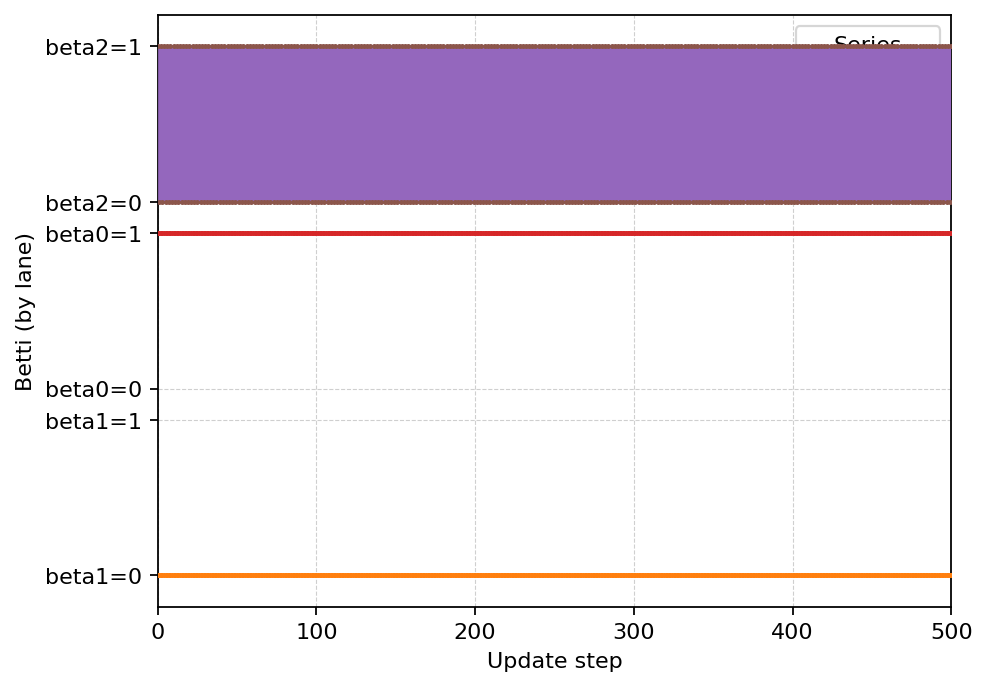}
    \caption{Betti numbers across the update sequence}
    \label{fig:p1_betti}
\end{subfigure}
\label{fig:p1}
\caption{}
\end{figure}

The latency series in Fig.~\ref{fig:p1_latency} is nearly flat between periodic recompressions (configured at 32 updates). Outside these events, the MMHM curve remains essentially constant, whereas baselines exhibit sustained oscillations over long stretches reflecting sensitivity to fill-in, pivot cascades, and global reduction orderings. These fluctuations increase variance and degrade amortized responsiveness relative to MMHM.

Homology verification in Fig.~\ref{fig:p1_betti} matches the specification. The complex stays connected and genus-0, with a single boundary component only in the open state; consequently \(\beta_0=1\) and \(\beta_1=0\) are invariant, and \(\beta_2\) alternates \(1\!\leftrightarrow\!0\) as the two-face window (with its shared edge) is opened and closed.

In sum, P1 isolates a strictly local edit on a triangulated sphere with bounded combinatorial deltas \((\Delta|E|=\pm 1,\ \Delta|F|=\pm 2,\ \text{fixed }|V|)\). Under this regime MMHM delivers stable, low-variance per-update cost between scheduled recompressions, with only rare threshold-triggered spikes, while baselines remain sensitive to global reduction effects.

\vspace{1em}

\subsection{P2: Locality Scaling under Topology-Preserving Refinement/Coarsening}

Let $K_0$ be a tetrahedral simplicial complex homeomorphic to a $3$-ball, with homology computed over a field (by default $\mathbb{F}=\mathbb{Z}_2$). Hence the ground-truth Betti profile is
\[
(\beta_0,\beta_1,\beta_2)=(1,0,0),
\]
and remains invariant throughout the benchmark. Each update acts on a small star-shaped region $R_t\subset \mathrm{int}(K)$ containing $s_t$ tetrahedra at time $t$, modifying only the local tessellation while preserving the PL homeomorphism type.

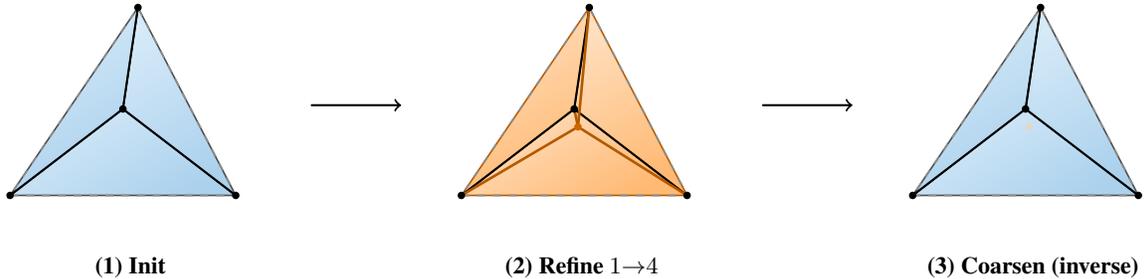
\begin{figure}[h]
\centering
\begin{tikzpicture}[scale=1.0, every node/.style={font=\small}]
  \definecolor{faceA}{RGB}{214,234,248}
  \definecolor{faceB}{RGB}{171,209,237}
  \tikzset{
    tface/.style={
      draw=black!70, line width=0.6pt,
      shading=axis, left color=faceA, right color=faceB, shading angle=30
    },
    tfaceHi/.style={
      draw=orange!70!black, line width=0.6pt,
      shading=axis, left color=orange!25, right color=orange!55, shading angle=30
    },
    edge/.style={draw=black, line width=0.8pt, line join=round, line cap=round},
    edgeStrong/.style={draw=orange!70!black, line width=1.0pt, line join=round, line cap=round},
    hidden/.style={draw=black!45, line width=0.7pt, dashed},
    vtx/.style={circle, fill=black, inner sep=1pt}
  }

  \def\Ax{0}   \def\Ay{0}
  \def\Bx{3}   \def\By{0}
  \def\Cx{1.7} \def\Cy{2.5}
  \def\Dx{1.5} \def\Dy{1.15}
  \pgfmathsetmacro{\Ex}{(\Ax+\Bx+\Cx+\Dx)/4}
  \pgfmathsetmacro{\Ey}{(\Ay+\By+\Cy+\Dy)/4}

  \begin{scope}[shift={(0,0)}]
    \fill[tface] (\Dx,\Dy)--(\Ax,\Ay)--(\Bx,\By)--cycle;
    \fill[tface] (\Dx,\Dy)--(\Bx,\By)--(\Cx,\Cy)--cycle;
    \fill[tface] (\Dx,\Dy)--(\Cx,\Cy)--(\Ax,\Ay)--cycle;
    \draw[hidden] (\Ax,\Ay)--(\Bx,\By)--(\Cx,\Cy)--cycle;
    \draw[edge] (\Dx,\Dy)--(\Ax,\Ay);
    \draw[edge] (\Dx,\Dy)--(\Bx,\By);
    \draw[edge] (\Dx,\Dy)--(\Cx,\Cy);
    \foreach \x/\y in {\Ax/\Ay,\Bx/\By,\Cx/\Cy,\Dx/\Dy}{\node[vtx] at (\x,\y) {};}
    \node[below] at (1.6,-0.7) {\textbf{(1) Init}};
  \end{scope}

  \draw[->,thick] (4.0,1.2) -- (5.2,1.2);

  \begin{scope}[shift={(6.0,0)}]
    \fill[tfaceHi] (\Dx,\Dy)--(\Ax,\Ay)--(\Bx,\By)--cycle;
    \fill[tfaceHi] (\Dx,\Dy)--(\Bx,\By)--(\Cx,\Cy)--cycle;
    \fill[tfaceHi] (\Dx,\Dy)--(\Cx,\Cy)--(\Ax,\Ay)--cycle;
    \draw[hidden] (\Ax,\Ay)--(\Bx,\By)--(\Cx,\Cy)--cycle;
    \draw[edge] (\Dx,\Dy)--(\Ax,\Ay);
    \draw[edge] (\Dx,\Dy)--(\Bx,\By);
    \draw[edge] (\Dx,\Dy)--(\Cx,\Cy);
    \node[vtx,fill=orange!80!black] (E) at (\Ex,\Ey) {};
    \foreach \x/\y in {\Ax/\Ay,\Bx/\By,\Cx/\Cy,\Dx/\Dy}{\draw[edgeStrong] (E) -- (\x,\y);}
    \foreach \x/\y in {\Ax/\Ay,\Bx/\By,\Cx/\Cy,\Dx/\Dy}{\node[vtx] at (\x,\y) {};}
    \node[below] at (1.6,-0.7) {\textbf{(2) Refine $1{\to}4$}};
  \end{scope}

  \draw[->,thick] (10.0,1.2) -- (11.2,1.2);

  \begin{scope}[shift={(12.0,0)}]
    \fill[tface] (\Dx,\Dy)--(\Ax,\Ay)--(\Bx,\By)--cycle;
    \fill[tface] (\Dx,\Dy)--(\Bx,\By)--(\Cx,\Cy)--cycle;
    \fill[tface] (\Dx,\Dy)--(\Cx,\Cy)--(\Ax,\Ay)--cycle;
    \draw[hidden] (\Ax,\Ay)--(\Bx,\By)--(\Cx,\Cy)--cycle;
    \draw[edge] (\Dx,\Dy)--(\Ax,\Ay);
    \draw[edge] (\Dx,\Dy)--(\Bx,\By);
    \draw[edge] (\Dx,\Dy)--(\Cx,\Cy);
    \node[vtx,fill=orange!40,opacity=0.45] at (\Ex,\Ey) {};
    \foreach \x/\y in {\Ax/\Ay,\Bx/\By,\Cx/\Cy,\Dx/\Dy}{\node[vtx] at (\x,\y) {};}
    \node[below] at (1.6,-0.7) {\textbf{(3) Coarsen (inverse)}};
  \end{scope}
\end{tikzpicture}
\caption{Three-step local cycle in region $R$}
\label{fig:tets_1to4_cycle}
\end{figure}

The update sequence $\{u_t\}_{t=1}^T$ alternates between a local refinement and its exact inverse:
\[
\textsc{Refine}(R_t):\ K \mapsto \mathrm{stellar}_{1\to4}(K;R_t),\qquad
\textsc{Coarsen}(R_t):\ K \mapsto \mathrm{inverse}(K;R_t)
\]
where $\mathrm{stellar}_{1\to4}$ applies tetrahedral $1{\to}4$ stellar subdivisions to all $\tau\in R_t$ (any auxiliary edge/face splits are confined to $R_t$), and $\mathrm{inverse}$ reverts precisely those local subdivisions via matched edge collapses/tet merges within $R_t$. Since $R_t\subset \mathrm{int}(K)$, the boundary $\partial K$ is unchanged up to subdivision and no boundary components are created or removed; connectivity is preserved. Consequently,
\[
\beta_0 \equiv 1,\qquad \beta_1 \equiv 0,\qquad \beta_2 \equiv 0
\]

Locality is parameterized by $s=\max_k s_k$, where $s_k$ counts the $k$-simplices whose incidences change during an update; in the reduced complex, the corresponding affected critical columns are $s_k^C$. No combinatorial gate is used in P2; the evaluation isolates the cost of localized rank maintenance in 3D. Periodic recompression is performed every $32$ steps, and threshold-triggered rebuilds are enabled but rare. In the intended regime with $s_k^C \ll n_k^C$, MMHM restricts elimination to the affected critical block, yielding a near-quadratic dependence in $s_k^C$, whereas baselines scale with the total critical size $|C|$.

\begin{figure}[!hb]
    \centering
    \includegraphics[width=1.0\textwidth]{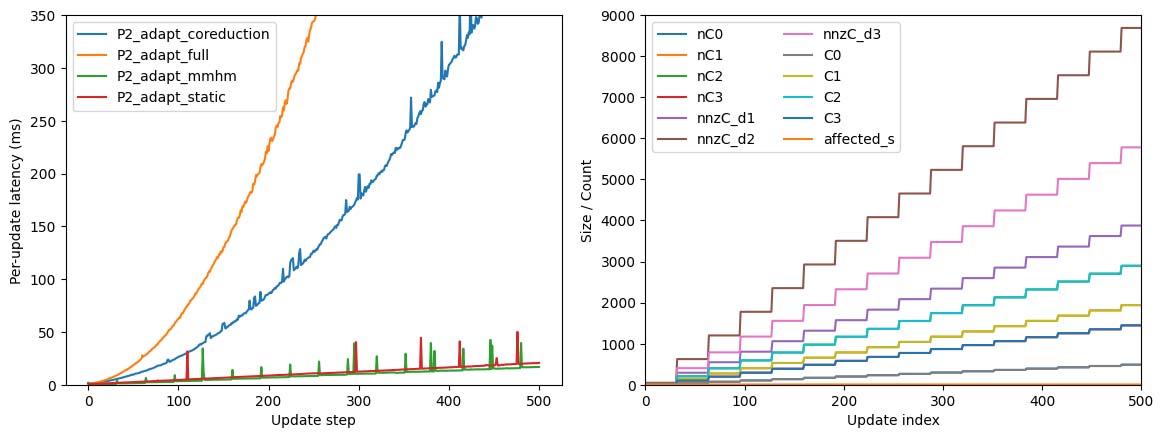}
    \caption{P2 benchmark. \emph{Left:} Per-update latency for P2 benchmark. 
\emph{Right:} MMHM critical size over time}
    \label{fig:p2_latency}
\end{figure}

Despite the absence of a gate and the fact that $s_k^C$ can temporarily approach $n_k^C$ between recompressions, MMHM maintains a small but consistent edge over Static--PH. The right panel explains MMHM’s effective problem size follows a staircase because elimination is confined to the affected critical block and global fill-in is suppressed. Static--PH benefits from strong one-shot Morse compression and cache-friendly streaming reductions, narrowing the gap. As locality increases or sequences lengthen, MMHM typically widens its advantage; the recompression overhead is fixed and amortized.

\begin{figure}[h]
    \centering
    \includegraphics[width=\textwidth]{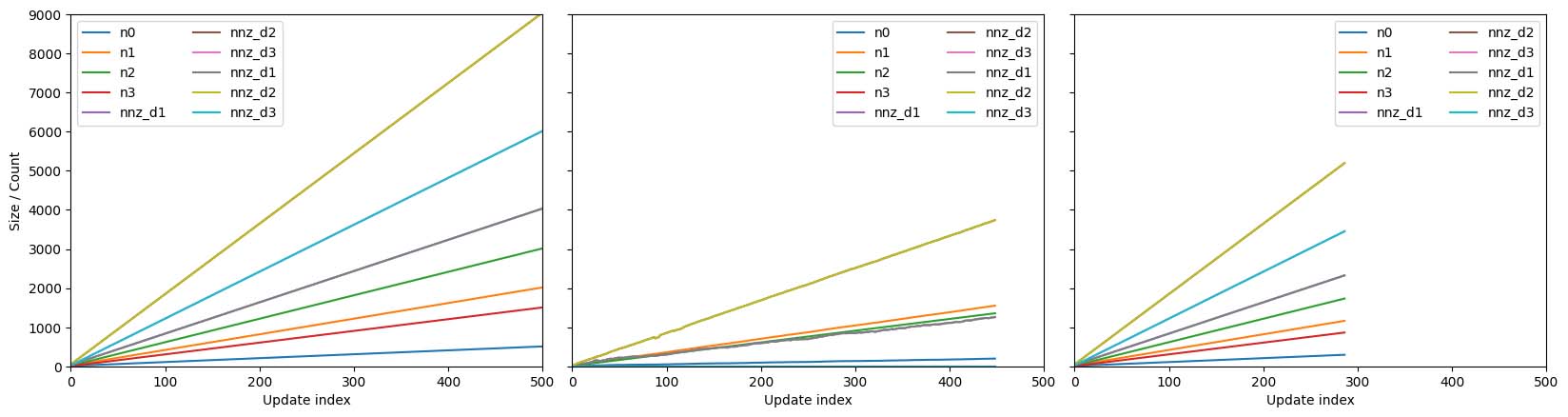}
    \caption{Size comparison of boundary data across the update sequence for P2 \emph{Left:} Static--PH. \emph{Middle:} Coreduction. \emph{Right:} Full recomputation.}
    \label{p2_size}
\end{figure}

As shown in Fig.~\ref{p2_size}, the size curves measured by the total number of critical simplices $|C|$ and the nonzeros $\mathrm{nnz}(\mathbf{B}_k)$ for global methods (Static--PH, Coreduction, Full) grow approximately linearly with the step index $t$. Each step applies a fixed-size $1{\to}4$ refinement within a small star, injecting $O(1)$ new simplices per step; thus $|C_t|=\Theta(t)$ and any pipeline that reprocesses the entire (critical) complex at every step naturally tracks this linear growth. On this benchmark, the Coreduction and Full recomputation baselines did not finish within the time budget: repeated global reductions over an increasingly dense neighborhood exhaust time and memory as elimination progressively behaves like a dense block. This behavior is consistent with the complexity envelope discussed earlier, where full reductions scale at least quadratically in matrix dimension and can approach cubic time under adverse fill-in; repeated $1{\to}4$ refinements near a fixed star realize precisely the conditions that increase $\mathrm{nnz}(\mathbf{B}_k)$ and amplify fill-in over time.

P2 confirms that, under PL-invariant 3D updates with localized changes, MMHM confines work to the affected critical block and achieves lower amortized latency than Static--PH. The staircase evolution of the effective critical size accounts for the observed margin, and the advantage typically grows with locality or sequence length while recompression cost remains amortized. This establishes a baseline for subsequent benchmarks that introduce gates and genuine topological events.

\subsection{P3: Dual-Forest Cavity/Duct Maintenance}\label{sec:p3}
The forest-shaped duct network bounds the number of faces incident to any port by a constant and prevents long $V$-paths of cancellations from propagating far from the edit site. Consequently, the number of affected critical $2$-cells and the fill-in during local reductions remain near-constant across the sequence. This setting stresses the ability of maintenance algorithms to exploit locality: methods whose per-update cost scales with global matrix size $|C|$ are disadvantaged relative to algorithms whose cost scales with the local footprint $s$.

\begin{figure}[!hb]
\centering
\begin{tikzpicture}[x=1cm,y=1cm,every node/.style={font=\small}]
\tikzset{
  shell/.style={line width=1.0pt},
  cavity/.style={draw=black, line width=0.8pt, fill=white},
  duct/.style={line width=1.2pt},
  portOpen/.style={draw=green!55!black, line width=1.2pt},
  portSealed/.style={draw=black, fill=gray!35, line width=0.8pt},
  locality/.style={draw=blue!60, dashed, line width=0.8pt},
  legend/.style={font=\footnotesize}
}

\node[anchor=west] at (-4.2,3.7) {\textbf{(A) Shell, cavities, ducts, ports}};

\def\R{3.2}
\draw[shell] (0,0) circle (\R);

\node[cavity, circle, inner sep=0pt, minimum size=1.10cm] (C1) at (-1.20, 0.80) {$C_1$};
\node[cavity, circle, inner sep=0pt, minimum size=1.10cm] (C2) at ( 1.05, 1.00) {$C_2$};
\node[cavity, circle, inner sep=0pt, minimum size=1.10cm] (C3) at ( 0.15,-1.35) {$C_3$};

\draw[duct] (C1.east) -- (C2.west);
\draw[duct] (C2.south west) -- (C3.north);

\coordinate (pOpen) at ({\R*cos(150)},{\R*sin(150)});
\coordinate (pC1mid) at ($ (C1)!0.55!(pOpen) $);
\draw[duct] (C1.north west) -- (pC1mid) -- (pOpen);

\draw[portOpen,fill=white] (pOpen) circle (0.18);
\node[legend,anchor=east] at ($(pOpen)+(-0.05,0.45)$) {open port};

\draw[locality] (pOpen) circle (0.70);
\node[legend,anchor=west,align=left] at ($(pOpen)+(0.78,0.00)$)
{local edit neighborhood\\(radius $\approx s$)};

\coordinate (pSeal) at ({\R*cos(-35)},{\R*sin(-35)});
\draw[duct] (C3.south) .. controls ($ (C3.south) + (0.5,-0.5) $) .. (pSeal);

\draw[portSealed] (pSeal) circle (0.18);
\node[legend,anchor=west] at ($(pSeal)+(0.28,0.00)$) {sealed port (membrane)};

\node[anchor=west] at (4.6,3.7) {\textbf{(B) Dual forest (cavities as nodes)}};

\node[circle,draw,inner sep=2pt] (v1) at (5.3, 1.6) {$C_1$};
\node[circle,draw,inner sep=2pt] (v2) at (6.7, 0.4) {$C_2$};
\node[circle,draw,inner sep=2pt] (v3) at (5.1,-0.9) {$C_3$};

\node[rectangle,draw,inner sep=2pt,fill=white] (ext) at (8.4,0.6) {Exterior};

\draw[line width=1.1pt] (v1) -- (v2);
\draw[line width=1.1pt] (v2) -- (v3);

\draw[portOpen] (v1) -- node[legend,midway,above=1pt] {open} (ext);

\draw[line width=1.0pt] (v3) -- ($(v3)!0.78!(ext)$);
\draw[line width=1.0pt] ($(v3)!0.78!(ext)$) -- (ext);
\draw[line width=1.6pt] ($(v3)!0.78!(ext) + (0,0.07)$) -- ($(v3)!0.78!(ext) + (0,-0.07)$);
\node[legend,anchor=west] at (8.6,-0.2) {sealed};

\end{tikzpicture}
\caption{}
\label{fig:p3_tikz}
\end{figure}

The complex is a triangulated $2$-manifold (“shell”) embedded in $\mathbb{R}^3$ that abstracts the effect of multiple interior cavities. Instead of explicitly constructing cavities and ducts, their topological effect is represented via small circular “ports” on the shell (each a constant-size triangulated disk). The putative duct network is treated through the dual graph of cavity components: vertices correspond to conceptual cavities and edges correspond to conceptual ducts. This network is constrained to be a forest (acyclic) at the level of abstraction, ensuring that local edits near port neighborhoods do not accumulate global handles on the shell. Figure~\ref{fig:p3_tikz} illustrates the setup: panel (A) depicts the shell with three representative port neighborhoods (labeled $C_1,C_2,C_3$ for reference), and two boundary ports—one \emph{open} and one \emph{sealed}. A dashed circle marks the local edit neighborhood of radius $\approx s$. For a genus-$0$ shell with $b$ open boundary ports, the expected Betti numbers are
$\beta_0=1,\ \beta_2=\mathbf{1}[b=0],\ \beta_1=\max\{0,b-1\}.$
Panel (B) shows the dual-graph \emph{abstraction}: cavity nodes linked by duct edges, plus an Exterior node; an edge to Exterior represents an open port, whereas a barred edge segment indicates a sealed port. The dual graph being acyclic emphasizes that updates do not induce long-range topological dependencies.

\begin{figure}[H]
\centering
\begin{tikzpicture}[x=1cm,y=1cm,every node/.style={font=\small}]
\tikzset{
  shell/.style={line width=1.0pt},
  cavity/.style={draw=black, line width=0.8pt, fill=white},
  duct/.style={line width=1.2pt},
  portOpen/.style={draw=green!55!black, line width=1.2pt},
  portSealed/.style={draw=black, fill=gray!35, line width=0.8pt},
  locality/.style={draw=blue!60, dashed, line width=0.8pt},
  legend/.style={font=\footnotesize}
}

\newcommand{\TitlePad}{0.25cm}
\newcommand{\SubPad}{0.28cm}

\def\R{2.2}
\def\thA{-15}
\def\thB{85}
\def\thP{25}
\def\radL{1.15}

\begin{scope}[shift={(-4.6,0)}, local bounding box=LBox]
  \draw[shell] (0,0) ++(\thA:\R) arc (\thA:\thB:\R);

  \coordinate (pOpen) at ({\R*cos(\thP)},{\R*sin(\thP)});

  \coordinate (qInL) at (-0.5,-0.6);
  \draw[duct] (qInL) .. controls (0.3,-0.2) .. (pOpen);

  \draw[portOpen,fill=white] (pOpen) circle (0.16);
  \draw[locality] (pOpen) circle (\radL);

    \coordinate (OpenLabel) at ($(pOpen)+(1,0.05)$);
    \node[legend,anchor=west] at (OpenLabel) {$-\,$constant triangles};
    
    \coordinate (triOpenA) at ($ (OpenLabel)+(-0.34,-0.02) $);
    \coordinate (triOpenB) at ($ (OpenLabel)+(-0.26,-0.12) $);
    \draw[line width=0.6pt] (triOpenA) -- ++(0.18,0) -- ++(-0.09,0.14) -- cycle;
    \draw[line width=0.6pt] (triOpenB) -- ++(0.18,0) -- ++(-0.09,0.14) -- cycle;

  \node[legend,anchor=north] at ($(pOpen)+(0,-\radL-0.5)$) {locality $\approx s$};
\end{scope}

\begin{scope}[shift={(4.6,0)}, local bounding box=RBox]
  \draw[shell] (0,0) ++(\thA:\R) arc (\thA:\thB:\R);

  \coordinate (pSeal) at ({\R*cos(\thP)},{\R*sin(\thP)});

  \coordinate (qInR) at (-0.4,-0.7);
  \draw[duct] (qInR) .. controls (0.35,-0.25) .. (pSeal);

  \draw[portSealed] (pSeal) circle (0.16);
  \draw[locality] (pSeal) circle (\radL);

    \coordinate (SealLabel) at ($(pSeal)+(1,0.05)$);
    \node[legend,anchor=west] at (SealLabel) {$+\,$constant triangles};
    
    \coordinate (triSealA) at ($ (SealLabel)+(-0.34,-0.02) $);
    \coordinate (triSealB) at ($ (SealLabel)+(-0.26,-0.12) $);
    \fill[gray!35] (triSealA) -- ++(0.18,0) -- ++(-0.09,0.14) -- cycle;
    \fill[gray!35] (triSealB) -- ++(0.18,0) -- ++(-0.09,0.14) -- cycle;

  \node[legend,anchor=north] at ($(pSeal)+(0,-\radL-0.5)$) {locality $\approx s$};
\end{scope}

\node[legend,anchor=south] at ($(LBox.north)+(0,\SubPad)$) {\textbf{Before: open}};
\node[legend,anchor=south] at ($(RBox.north)+(0,\SubPad)$) {\textbf{After: sealed}};

\node[anchor=south] at ($(current bounding box.north)+(0,\TitlePad)$)
{\textbf{Local topological edit at a port (open $\leftrightarrow$ seal)}};

\end{tikzpicture}
\label{fig:p3_local_toggle}
\end{figure}

All edits are confined to a local neighborhood of exactly one port. Two atomic topological actions are allowed at a port: \emph{open} (remove a small membrane patch by deleting a constant number of triangles) and \emph{seal} (insert the membrane patch by adding the same pattern of triangles). Local re-meshing ($2$–$2$ edge flips in 2D charts; small Pachner moves in 3D charts) is permitted only to regularize the mesh without changing the port’s open/closed state or merging/splitting boundary loops. No other global surgery is allowed. The locality parameter $s$ denotes the maximum number of simplices touched by one edit; in benchmarks, $s$ is small and fixed across the sequence. Each shell component remains connected ($\beta_0=1$). Within a component, when all ports are sealed the surface is closed and orientable, hence $\beta_2=1$ and $\beta_1=2g$, where $g$ is the (fixed) genus of that component (typically $g=0$ in the benchmark construction).

Thus, toggling a single port changes $b$ by $\pm 1$ and adjusts $(\beta_1,\beta_2)$ accordingly in a predictable, local manner. The forest constraint prevents port operations from creating global cycles that would alter $g$.

\begin{table}[!hb]
\centering
\caption{Mean per-update runtime on P3}
\label{tab:p3_latency}
\begin{tabular}{lccc}
\toprule
Method & Amortized (ms/update) & Mean step time (ms) & Throughput (updates/s) \\
\midrule
MMHM            & 28.13 & 17.18 & 35.54 \\
Coreduction     & 356.28 & 350.97 & 2.81 \\
Full-Recompute  & 409.89 & 404.47 & 2.44 \\
\bottomrule
\end{tabular}
\end{table}

Table~\ref{tab:p3_latency} reports per-update runtimes. MMHM attains an amortized $28.13\,\mathrm{ms}$ per update and a mean step time of $17.18\,\mathrm{ms}$, whereas Coreduction and Full-Recompute are in the $350$--$410\,\mathrm{ms}$ range. In relative terms, the amortized latency reduction of MMHM is $12.7\times$ over Coreduction and $14.6\times$ over Full-Recompute. On mean step time the gap is even larger: $20.4\times$ and $23.5\times$, respectively. Throughput follows the same pattern: $36.95$ updates/s for MMHM versus $2.89$ and $2.50$ for Coreduction and Full-Recompute, i.e., about $12.8\times$ and $14.8\times$ higher.

\begin{figure}[H]
\centering
\begin{subfigure}[b]{0.47\textwidth}
    \centering
    \includegraphics[width=\textwidth]{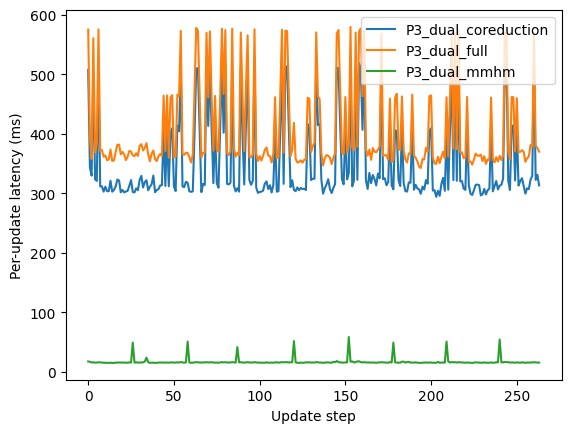}
    \caption{Per-update latency for P3 benchmark.}
    \label{fig:p3_latency}
\end{subfigure}
\hfill
\begin{subfigure}[b]{0.47\textwidth}
    \centering
    \includegraphics[width=\textwidth]{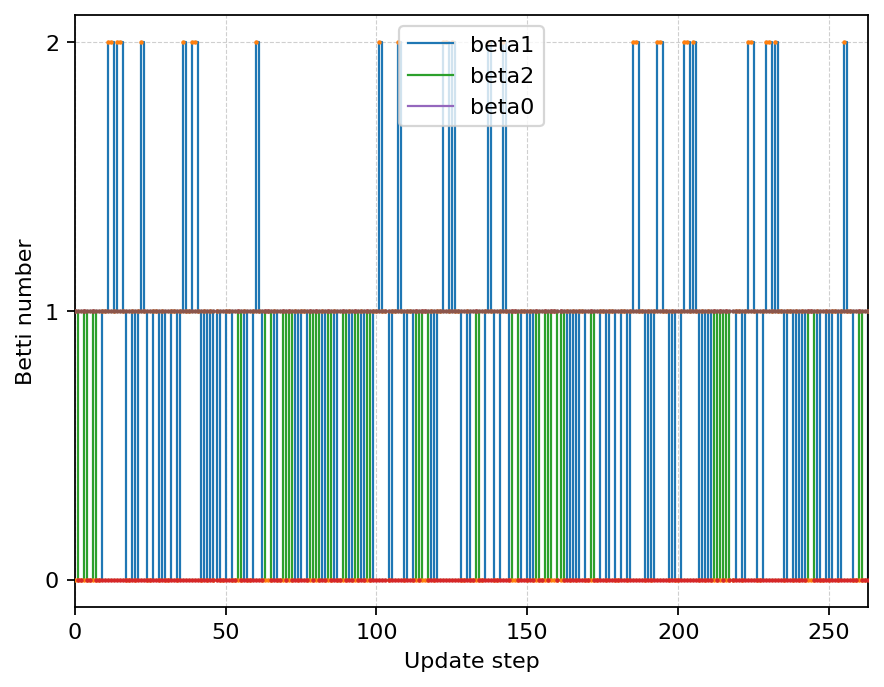}
    \caption{Betti numbers across the update sequence}
    \label{fig:p3_betti}
\end{subfigure}
\label{fig:p3}
\caption{}
\end{figure}

Figure~\ref{fig:p3_latency} shows the temporal structure of these costs. The MMHM curve remains near a low baseline with shallow, regular undulations, consistent with periodic housekeeping/rebuilds amortized over the sequence. In contrast, both Full-Recompute and Coreduction fluctuate widely at a much higher scale, reflecting their sensitivity to local mesh variability even when each edit touches only a fixed-size neighborhood. The stability of MMHM is notable: despite minor periodic ripples, the latency envelope is essentially flat relative to the baselines’ variance. The difference between ``Amortized (ms/update)'' and ``Mean step time (ms)'' in Table~\ref{tab:p3_latency} is informative. The amortized quantity includes occasional non-regular work (e.g., scheduled rebuilds and initialization) divided across all updates, while the mean step time averages the per-edit costs observed on regular steps. For MMHM, the amortized value exceeds the mean step value, indicating that infrequent heavy operations exist but are rare enough that regular steps remain fast; for the baselines, the two columns are close because almost every step is expensive. Consequently, throughput aligns more closely with the inverse of the amortized cost (effective end-to-end rate) than with the inverse of the mean step time, which is consistent with the shapes seen in Fig.~\ref{fig:p3_latency}.

Fig.~\ref{fig:p3_betti} confirms that all methods track the same homological trajectory throughout the sequence. In particular, $\beta_0$ remains identically $1$ (the shell never disconnects), openings and sealings toggle boundary loops so that $\beta_1$ changes in unit steps aligned with those events, and $\beta_2$ equals $1$ only in the temporarily fully sealed state and drops to $0$ whenever at least one port is open. Because each edit modifies only a constant-size topological disk and local re-meshing does not merge or split boundary loops, no method can introduce a handle in this benchmark. The Betti traces, therefore, show that identical curves across methods indicate that the implementations preserve the intended surface topology under strictly local edits.

MMHM achieves an order-of-magnitude speedup on P3 while retaining the same homological behavior. The latency profile remains near a low baseline with mild periodic ripples attributable to scheduled housekeeping, whereas the baselines fluctuate at a much higher scale. Taken together with the matching Betti trajectories, these results show that MMHM sustains high update rates without sacrificing topological correctness.

\section{Conclusion}
This work introduced MMHM, a practical framework for dynamic maintenance of low-dimensional simplicial homology over $\mathbb{Z}_2$ that combines discrete Morse compression to a critical complex, localized boundary updates with cached pivot ownership, and conservative topological gates to bypass linear algebra when invariants are decidable combinatorially; a periodic recompression policy preserves sparsity and keeps the critical representation synchronized with the evolving complex. Analysis yields a linear–to–quadratic envelope for localized reduction. Empirically, near-linear scaling is observed in locality-driven workloads such as topology-preserving refinement/coarsening, while minimal topological perturbations on genus-$0$ triangulations and cavity/duct updates in 3D meshes exhibit flat latency traces punctuated by isolated recompression spikes. Overall, the framework bridges theory and systems, translating chain-homotopy equivalence into concrete savings in memory movement and elimination work, offers a drop-in upgrade for topology pipelines by turning costly rebuilds into fast, exact local updates, and reframes dynamic homology as a locality, and bounded maintenance task that provides an exact alternative to global recomputation for evolving meshes and complexes.

\bibliographystyle{unsrtnat}
\bibliography{template}

\begin{thebibliography}{15}
\providecommand{\natexlab}[1]{#1}
\providecommand{\url}[1]{\texttt{#1}}
\expandafter\ifx\csname urlstyle\endcsname\relax
  \providecommand{\doi}[1]{doi: #1}\else
  \providecommand{\doi}{doi: \begingroup \urlstyle{rm}\Url}\fi

\bibitem[Edelsbrunner et~al.(2002)Edelsbrunner, Letscher, and Zomorodian]{edelsbrunner2002topological}
Herbert Edelsbrunner, David Letscher, and Afra Zomorodian.
\newblock Topological persistence and simplification.
\newblock \emph{Discrete \& Computational Geometry}, 28\penalty0 (4):\penalty0 511--533, 2002.
\newblock \doi{10.1007/s00454-002-2885-2}.

\bibitem[Hatcher(2002)]{hatcher2002algebraic}
Allen Hatcher.
\newblock \emph{Algebraic Topology}.
\newblock Cambridge University Press, 2002.

\bibitem[Carlsson(2009)]{carlsson2009topology}
Gunnar Carlsson.
\newblock Topology and data.
\newblock \emph{Bulletin of the American Mathematical Society}, 46\penalty0 (2):\penalty0 255--308, 2009.
\newblock \doi{10.1090/S0273-0979-09-01249-X}.

\bibitem[Zomorodian and Carlsson(2005)]{zomorodian2005computing}
Afra Zomorodian and Gunnar Carlsson.
\newblock Computing persistent homology.
\newblock \emph{Discrete \& Computational Geometry}, 33\penalty0 (2):\penalty0 249--274, 2005.
\newblock \doi{10.1007/s00454-004-1146-y}.

\bibitem[Munkres(1984)]{munkres1984elements}
James~R Munkres.
\newblock \emph{Elements of Algebraic Topology}.
\newblock Addison-Wesley, 1984.

\bibitem[Bauer(2021)]{bauer2021ripser}
Ulrich Bauer.
\newblock Ripser: efficient computation of vietoris--rips persistence barcodes.
\newblock \emph{Journal of Applied and Computational Topology}, 5\penalty0 (3):\penalty0 391--423, 2021.
\newblock \doi{10.1007/s41468-021-00071-5}.

\bibitem[Pachner(1991)]{pachner1991pl}
Udo Pachner.
\newblock Pl homeomorphic manifolds are equivalent by elementary shellings.
\newblock \emph{European Journal of Combinatorics}, 12\penalty0 (2):\penalty0 129--145, 1991.
\newblock \doi{10.1016/S0195-6698(13)80080-7}.

\bibitem[Forman(2002)]{forman2002user}
Robin Forman.
\newblock A user's guide to discrete morse theory.
\newblock \emph{S{\'e}minaire Lotharingien de Combinatoire}, 48:\penalty0 B48c, 2002.

\bibitem[Edelsbrunner and Harer(2010)]{edelsbrunner2010computational}
Herbert Edelsbrunner and John~L. Harer.
\newblock \emph{Computational Topology: An Introduction}.
\newblock American Mathematical Society, 2010.
\newblock ISBN 978-0-8218-4925-5.
\newblock \doi{10.1007/978-3-540-33259-6_7}.
\newblock URL \url{http://www.ams.org/bookstore-getitem/item=MBK-69}.

\bibitem[Mrozek and Batko(2009)]{MrozekBatko2009Coreduction}
Marian Mrozek and Bogdan Batko.
\newblock Coreduction homology algorithm.
\newblock \emph{Discrete \& Computational Geometry}, 41\penalty0 (1):\penalty0 96--118, 2009.
\newblock \doi{10.1007/s00454-008-9073-y}.

\bibitem[D{\l}otko et~al.(2011)D{\l}otko, Kaczynski, Mrozek, and Wanner]{DlotkoEtAl2011CoreductionCW}
Pawe{\l} D{\l}otko, Tomasz Kaczynski, Marian Mrozek, and Thomas Wanner.
\newblock Coreduction homology algorithm for regular cw-complexes.
\newblock \emph{Discrete \& Computational Geometry}, 46\penalty0 (2):\penalty0 361--388, 2011.
\newblock \doi{10.1007/s00454-010-9303-y}.

\bibitem[Harker et~al.(2014)Harker, Mischaikow, Mrozek, and Nanda]{HarkerEtAl2013FoCM}
Shaun Harker, Konstantin Mischaikow, Marian Mrozek, and Vidit Nanda.
\newblock Discrete morse theoretic algorithms for computing homology of complexes and maps.
\newblock \emph{Foundations of Computational Mathematics}, 14\penalty0 (1):\penalty0 151--184, 2014.
\newblock \doi{10.1007/s10208-013-9145-0}.

\bibitem[Mrozek and Wanner(2010)]{MrozekWanner2010CoreductionInclusionsPH}
Marian Mrozek and Thomas Wanner.
\newblock Coreduction homology algorithm for inclusions and persistent homology.
\newblock \emph{Computers \& Mathematics with Applications}, 60\penalty0 (10):\penalty0 2812--2833, 2010.
\newblock \doi{10.1016/j.camwa.2010.08.063}.

\bibitem[Mischaikow and Nanda(2013)]{mischaikow2013morse}
Konstantin Mischaikow and Vidit Nanda.
\newblock Morse theory for filtrations and efficient computation of persistent homology.
\newblock \emph{Discrete \& Computational Geometry}, 50\penalty0 (2):\penalty0 330--353, 2013.
\newblock \doi{10.1007/s00454-013-9529-6}.

\bibitem[Lewiner et~al.(2003)Lewiner, Lopes, and Tavares]{lewiner2003optimal}
Thomas Lewiner, H{\'e}lio Lopes, and Geovan Tavares.
\newblock Optimal discrete morse functions for 2-manifolds.
\newblock \emph{Computational Geometry}, 26\penalty0 (3):\penalty0 221--233, 2003.
\newblock \doi{10.1016/S0925-7721(03)00004-9}.

\end{thebibliography}

\end{document}